\documentclass[acmlarge]{acmart}
\AtBeginDocument{%
  }

\makeatletter
\newcommand{\myconfshort}{\acmConference@shortname}
\newcommand{\myconffull}{\acmConference@name}
\newcommand{\myconfdate}{\acmConference@date}
\newcommand{\myconfloc}{\acmConference@venue}
\AtBeginDocument{
  \fancypagestyle{firstpagestyle}{
    \fancyhead{}%
    \fancyfoot[C]{}%
  }
  \fancyhf{}
  \fancyhead[LO]{\@headfootfont\shorttitle}%
  \fancyhead[RE]{\@headfootfont\@shortauthors}%
  \fancyhead[LE]{\@headfootfont\footnotesize \myconfshort, \myconfdate, \myconfloc}%
  \fancyhead[RO]{\@headfootfont\footnotesize \myconfshort, \myconfdate, \myconfloc}%
  \fancyfoot[C]{}%
}
\makeatother
\setcopyright{acmlicensed}
\copyrightyear{2026}
\acmYear{2026}
\setcopyright{cc}
\setcctype{by}
\acmConference[FAccT '26]{The 2026 ACM Conference on Fairness, Accountability, and Transparency}{June 25--28, 2026}{Montreal, QC, Canada}
\acmBooktitle{The 2026 ACM Conference on Fairness, Accountability, and Transparency (FAccT '26), June 25--28, 2026, Montreal, QC, Canada}
\acmDOI{10.1145/3805689.3812277}
\acmISBN{979-8-4007-2596-8/2026/06}

\usepackage{booktabs}
\usepackage{xspace}
\usepackage{colortbl}
\usepackage{tikz}
\usepackage{etoolbox} %
\usepackage{subcaption}
\usepackage{soul} %
\usepackage{placeins} %
\usepackage{tabularx}
\usepackage{array}
\usepackage{makecell}

\newif\ifdraft
\drafttrue            %
\newif\ifshowcomments %
\showcommentstrue     %

\definecolor{SpringGreen}{RGB}{0,255,127}
\definecolor{SkyBlue}{RGB}{135,206,235}
\definecolor{Lavender}{RGB}{230,230,250}
\definecolor{BrickRed}{RGB}{203,65,84}
\definecolor{TealBlue}{RGB}{54,117,136}
\definecolor{Purple}{RGB}{128,0,128}

\newcommand{\greenbox}[1]{#1}

\newcommand{\purplebox}[1]{#1}
\newcommand{\bluebox}[1]{#1}

\newcolumntype{Y}{>{\raggedright\arraybackslash}X}
\newcolumntype{C}{>{\centering\arraybackslash}X}

\usepackage[most]{tcolorbox}
\definecolor{takeawaycolor}{HTML}{F4EBDD}
\newcommand{\takeaway}[1]{\textbf{#1}}
\definecolor{takeawaycontrastcolor}{HTML}{F4EBDD}
\newcommand{\takeawaycontrast}[1]{\textbf{#1}}

\newcommand{\q}[1]{``#1''}
\newcommand{\totalparticipants}{24\xspace}

\usepackage{tabularx}
\usepackage{booktabs}
\usepackage{multirow}

\definecolor{theme1}{RGB}{230, 245, 255}  %
\definecolor{theme2}{RGB}{255, 245, 230}  %
\definecolor{theme3}{RGB}{245, 255, 230}  %
\definecolor{theme4}{RGB}{255, 230, 245}  %

\definecolor{theme1dark}{RGB}{180, 210, 230}
\definecolor{theme2dark}{RGB}{230, 210, 180}
\definecolor{theme3dark}{RGB}{210, 230, 180}
\definecolor{theme4dark}{RGB}{230, 180, 210}

\newcolumntype{M}[1]{>{\raggedright\arraybackslash}m{#1}}

\usepackage{array, longtable, multirow, booktabs}
\usepackage{graphicx} %

\definecolor{brandGreen}{RGB}{214,239,211}
\definecolor{brandPurple}{RGB}{234, 220, 248}
\definecolor{brandBlue}{RGB}{214,229,239}
\definecolor{brandYellow}{RGB}{253,246,197}

\renewcommand{\arraystretch}{1.2}
\usepackage{csquotes}

\newcommand{\tightenum}[1]{%
\vspace{-0.25\baselineskip}
\begin{enumerate}[leftmargin=*, nosep, topsep=0pt, itemsep=0pt, parsep=0pt]
#1
\end{enumerate}
\vspace{-0.75\baselineskip}
}
\definecolor{PaperBg}{HTML}{F7F6F3}
\definecolor{PaperRule}{HTML}{B8B1A3}
\definecolor{PaperText}{HTML}{222222}
\definecolor{BP}{HTML}{E6F0E9}
\definecolor{HW}{HTML}{E5EEF7}
\definecolor{CL}{HTML}{EEE7F2}
\usepackage{ragged2e}
\usepackage{enumitem}
\usepackage{makecell}
\definecolor{GovRecBg}{HTML}{F4EBDD}
\definecolor{GovRecFrame}{HTML}{8A6A3D}
\newtcolorbox{govrecbox}{
  boxrule=0.5pt,
  arc=1mm,
  colback=GovRecBg,
  colframe=GovRecFrame,
  coltext=PaperText,
  boxsep=0pt,
  left=5pt,
  right=5pt,
  top=3pt,
  bottom=3pt,
  before skip=4pt,
  after skip=4pt
}

\newcommand{\recommendation}[1]{%
  \begin{govrecbox}
    \textbf{Design artifacts}\par#1
  \end{govrecbox}
}

\usepackage{xparse}

\NewDocumentCommand{\theme}{m m o}{%
  \subsubsection{#1}%
  #2%
  \IfValueT{#3}{\recommendation{#3}}%
}

\colorlet{BPFrame}{PaperRule}
\colorlet{BPAccent}{green!35!black}

\colorlet{HWFrame}{PaperRule}
\colorlet{HWAccent}{blue!45!black}

\colorlet{CLFrame}{PaperRule}
\colorlet{CLAccent}{violet!35!black}

\tcbset{
  pqnotitle/.style={
    coltext=PaperText,
    boxrule=0.4pt,
    arc=1mm,
    left=2.5mm,right=2.5mm,top=1.8mm,bottom=1.8mm,
    before skip=5pt, after skip=5pt,
  }
}

\newtcolorbox{participantquoteBP}[1][]{%
  pqnotitle,
  colback=white,
  borderline west={2pt}{0pt}{BPAccent},
  #1
}

\newtcolorbox{participantquoteHW}[1][]{%
  pqnotitle,
  colback=white,
  borderline west={2pt}{0pt}{HWAccent},
  #1
}

\newtcolorbox{participantquoteCL}[1][]{%
  pqnotitle,
  colback=white,
  borderline west={2pt}{0pt}{CLAccent},
  #1
}

\newtcolorbox{participantquoteBPpair}[1][]{%
  pqnotitle,
  colback=white,
  borderline west={2pt}{0pt}{BPAccent},
  sidebyside,
  sidebyside gap=10pt,
  segmentation style={dashed, line width=0.5pt, draw=PaperRule},
  left=2.5mm, right=2.5mm,
  #1
}
\newcommand{\panelcaption}[1]{\textbf{\small #1}\par\smallskip}

\definecolor{RevBlue}{HTML}{00446B}
\definecolor{RevOrange}{HTML}{6E2E00}

\newcommand{\quoteattrib}[1]{\par\smallskip\hfill{\color{black}--- #1}}

\begin{document}

\title[Uncovering Trustworthiness Ideals in AI-powered Peripartum Information Seeking]{``Where is this coming from?'' Uncovering Trustworthiness Ideals in AI-powered Peripartum Information Seeking}

\author{Vaibhav Balloli}
\affiliation{\institution{University of Michigan}\city{Ann Arbor}\country{USA}}
\email{vballoli@umich.edu}
\author{Julia Erickson}
\affiliation{\institution{University of Michigan}\city{Ann Arbor}\country{USA}}
\email{juliaer@med.umich.edu}

\author{Xinyi Li}
\affiliation{\institution{University of Michigan}\city{Ann Arbor}\country{USA}}
\email{xinyiade@umich.edu}

\author{Erin MacMurray van Liemt}
\affiliation{%
  \institution{Google Research}
  \city{Los Angeles}
  \country{USA}
}
\email{evanliemt@google.com}

\author{Alex Peahl}
\affiliation{\institution{University of Michigan}\city{Ann Arbor}\country{USA}}
\email{alexfrie@med.umich.edu}

\author{Elizabeth Bondi-Kelly}
\affiliation{\institution{University of Michigan}\city{Ann Arbor}\country{USA}}
\email{ecbk@umich.edu}

\renewcommand{\shortauthors}{Balloli et al.}
\setcounter{tocdepth}{3}

\begin{abstract}
AI-powered tools increasingly promise to fill information gaps in health, especially in domains like maternal and reproductive health that demand timely, accurate, and actionable information. This is extremely important, as the United States leads peer nations in preventable deaths, with stark racial disparities. However, current AI and NLP-powered systems aim to improve access to vetted maternal health information by routing user queries to a factual response while under-specifying the socio-technical governance structures that shape trust, use, and harm in practice. We report findings from four synchronous focus groups ($n=24$) with three stakeholder groups central to peripartum information support: birthing people, clinicians, and health workers (e.g., doulas, social workers, community health workers) exploring topics around information seeking, experience with current clinical infrastructure, misinformation, and an AI-enabled factual answering tool design probe. Our inductive analysis surfaces a central finding: in high-stakes health contexts shaped by historical inequities, trustworthiness must be inspectable and not asserted. While stakeholders diverge on what makes information credible, they converge on the need for transparency, recourse, and ecosystem complementarity. Based on the discussions, we identify four themes and governance requirements: (1) support for social and identity-based sensemaking, (2) pluralistic verification practices, (3) inspectable governance with recourse mechanisms, and (4) ecosystem-aware integration that avoids shifting burden. Building on these findings, we propose design artifacts that are mistrust-aware and promote principled governance mechanisms for transparent, pluralistic AI systems. Finally, we discuss the implications of our findings for expanding human-AI evaluations and improving the transparency of deployed AI systems.
\end{abstract}

\ccsdesc[500]{Applied computing~Health care information systems}
\ccsdesc[500]{Applied computing~Consumer health}
\ccsdesc[300]{Social and professional topics~Medical technologies}

\keywords{health, reproductive health, maternal health, focus groups, online health information seeking, social media, human-AI interaction}

\received{20 February 2007}
\received[revised]{12 March 2009}

\begin{teaserfigure}
\centering
    \includegraphics[width=0.75\textwidth]{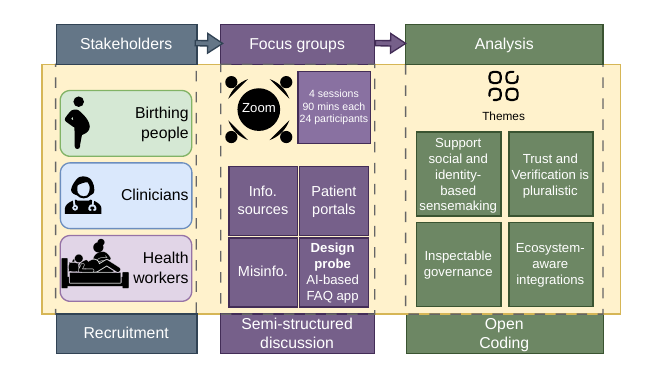}
    \caption{Overview of our study with three stages. Stage 1: relevant stakeholder recruitment, Stage 2: semi-structured, synchronous focus groups exploring four topics, Stage 3: Open coding and deriving insights.}
    \label{fig:teaser}
    \Description{Overview of our study with four stages. Stage 1: Focus groups with stakeholders, Stage 2: Analyzing data, Stage 3: Deriving themes, and Stage 4: Connecting themes and deriving insights.}
\end{teaserfigure}
  
 \maketitle

\section{Introduction}
\label{sec:intro}

The United States experiences higher rates of maternal death than peer countries, with stark inequities by race and socio-economic status~\cite{chinn2020maternal}. Recognizing complications during the pregnancy and postpartum period often depends on timely access to actionable information and responsive support~\cite{cdcprevention,acogredesign}, yet access to care is uneven. For example, 3.5\% of birthing people live in maternity care deserts, counties that lack key maternity care resources~\cite{maternitydesert}. Modern policy and infrastructure shifts have expanded reliance on Electronic Health Records (EHRs) and patient portals for patient-clinician communication~\cite{peahl2020patient}, with the promise to improve information access. However, clinicians spend substantial time in the EHR~\cite{rotenstein2021differences}, contributing to burden and burnout~\cite{adler2020electronic}. When clinical channels are slow, limited, or hard to navigate, many patients turn to faster alternatives such as social media, blogs, books, and trusted communities, despite recognizing the risk of misinformation.

These dynamics have motivated interventions ranging from health literacy and education~\cite{levy2016health,madewell2022prioritizing,peahl2021michigan} to conversational agents and information tools~\cite{nguyen2024rosie,jacaranda,ramjee2024ashabot,10.1145/2702123.2702427}. Active systems like Rosie~\cite{nguyen2024rosie} and Jacaranda~\cite{jacaranda} use state-of-the-art NLP and AI techniques to retrieve vetted documents and provide factual answers to maternal health questions. These approaches demonstrate great performance, yet in high-stakes domains like reproductive health, where trustworthiness is shaped by a complex social history and ongoing structural inequalities, \textbf{accurate information is necessary but not sufficient}. These concerns are also echoed in works like~\cite{antoniak2024nlp}, which emphasizes that systems and researchers should optimize for experiential outcomes, as opposed to a single interaction. Current AI-enabled information tools still rarely specify how governance for such systems should be designed: who defines what counts as vetted, how cultural and historical sensitivity is accounted for, how harms are detected and addressed, how accountability and recourse are provided when systems fail, and how these services integrate with clinical and community ecosystems without shifting burdens. Previous works~\cite{raji2020accountability,tang2024failurecards} have studied some of them and provided frameworks to better audit and report, yet open questions remain on how multiple stakeholders perceive them in a high-stakes domain like reproductive health. \phantomsection\label{par:intro2}

To examine this, we adopt a human-centered approach~\cite{bondi2021envisioning,birhane2022power} to understand the socio-technical aspects of AI-powered information support and uncover artifacts for designing accountable, trustworthy applications that can be embedded within clinical and community ecosystems, rather than as a standalone system. To ground design in lived experience and operational realities, we conducted synchronous focus groups with three stakeholder groups central to peripartum information seeking: birthing people, clinicians (e.g. Obstetrician-Gynecologists (OB/GYNs), Certified Nurse Midwives (CNMs), etc.), and health workers (e.g., doulas, social workers, community health workers). We explore the current information-seeking landscape, experiences with current medical systems, misinformation, and governance preferences around an AI-powered FAQ tool. With this multi-stakeholder design, we aimed to discover both shared and conflicting experiences, both within and across the stakeholder groups, all of whom are operating under the same healthcare system. \phantomsection\label{par:intro3}

\paragraph{Contributions.}
We position our contributions at the intersection of information seeking and governance of AI-enabled information systems, through the lens of reproductive health. Our focus is to characterize missing socio-technical requirements and to inform AI-based interventions aimed to bridge information gaps without shifting burdens or widening inequities. Our central finding is that \textbf{trustworthiness must be inspectable and not asserted}. Across stakeholder groups, participants reframed \q{good information} from factuality-only to actionable and context-aware information that can be safely integrated into existing care pathways. From our analysis, we uncover design implications around four themes: (1) support for social and identity-based sensemaking, (2) pluralistic verification practices, (3) inspectable governance with recourse mechanisms, and (4) ecosystem-aware integration that avoids shifting burden. Throughout our analyses, we examine fairness questions about whose experiences are centered and how verification works in communities with unequal trust in institutions, as well as accountability of these systems through provenance, recourse, and inspectable governance. \phantomsection\label{par:contributions}

\section{Background and Related Work}
\label{sec:background}
In this section, we first provide definitions of terms used throughout this paper. Then, we present related work on information seeking, misinformation, and systems in health that we build upon throughout the paper. Throughout this section, we highlight structural contexts and racial disparities that have been documented in reproductive health, and current AI systems with socio-technical requirements beyond accuracy.

\subsection{Definitions}
\label{subsec:definitions}

We adopt the following definitions throughout our study:
\begin{itemize}
    \item \textbf{Reproductive health:} We collectively refer to sexual and maternal health as reproductive health. Example topics include conception, pregnancy, and postpartum. 
    \item \textbf{Peripartum period:} \q{\textit{The period of time including before (prepregnancy), during, and after pregnancy (postpartum).}}\cite[p.~1586]{khan2023pregnancy}.
    \item \textbf{Obstetric care provider:} \q{\textit{A health care professional who specializes in pregnancy, labor, and delivery. These professionals include obstetrician-gynecologists (OB/GYNs), certified nurse-midwives (CNMs), maternal-fetal medicine specialists (MFMs), and family medicine doctors with experience in maternal care.}}~\cite[Under \q{Obstetric care provider}]{definition}
    \item \textbf{Health workers:} Professionals trained to support birthing people before birth, through labor, and after the birth of the baby, including emotional, informational, and physical support (e.g., doulas, social workers, community health workers). Specific role definitions can be found here~\cite[Under \q{Doula}, \q{social workers}]{definition}
    \item \textbf{Mistrust:} \q{\textit{A general sense of unease or suspicion toward someone or something that is predicated either on the notion that the provider or healthcare entity may not act in the patient’s best interest, and they may actively work against the patient. Mistrust may originate from distinct historical experiences linked to group identity, personal experience, vicarious experiences, and oral histories.}}\cite[p.~443]{griffith2021using}
    \item \textbf{Distrust:} \q{\textit{Distrusting beliefs of a patient may include $\geq$1 of the following components: competence, benevolence, integrity, predictability, and assurance.}}\cite[p.~443]{griffith2021using}
    \item \textbf{Pluralism:} Here, we define pluralism in the context that there can be multiple legitimate ways for stakeholders to navigate reproductive health information and no single method is inherently better, that is, similar to foundational value pluralism, \q{there is no one value that subsumes all other values, no one property of goodness, and no overarching principle of action}\cite{sep-value-pluralism}.
\end{itemize}

\subsection{Multi-faceted Information Seeking in Health}
\label{subsec:info_seeking}

Health information seeking encompasses multiple channels and motivations: digital platforms support understanding diagnoses~\cite{zuccon2015diagnose}, accessing others' experiences~\cite{eysenbach2004health,augustaitis2021online}, and obtaining health advice~\cite{johnson2006neo}. Other motivations include convenience, dissatisfaction with prior clinical explanations, and seeking recommendations and opinions~\cite{de2014seeking}. Interpersonal relationships like families, peers, and community networks have also been often stated to provide trusted guidance and experiential knowledge, especially among immigrant communities where these relationships are cultivated through community organizations and religious activities~\cite{BARNES2019493}. Structured information sources, including books, guidelines, brochures, and prenatal courses, complement this landscape~\cite{vogels2022sources}.

AI-based systems are increasingly a source of information in health. A key implication for information retrieval systems is that health information seeking is frequently multi-faceted, extending beyond fact retrieval; people seek reassurance, normalization, and strategies for deciding what to do next~\cite{turnbull2019conceptual}, influenced by their lived experiences and care infrastructure around them~\cite{10.1145/3491102.3501950}. These dynamics matter for the trustworthiness of these systems because they make clear that the harms of AI-mediated support are not limited to incorrect statements. Systems also shape what is treated as normal, and whose experiences are centered or ignored~\cite{10.1145/3544548.3581475}. For example, ~\cite{veinot2018good,zagloul2024review} document how current health systems have led to differential access and experiences based on socio-economic factors and religion, respectively. Hence, we aim to understand and articulate governance requirements for AI-based peripartum information systems that account for the underlying socio-technical factors: how knowledge is accessed and contextualized, how values and lived experience are represented, how uncertainties are handled, and what forms of accountability, recourse, and escalation are available when users need more than an answer.

\subsection{Misinformation and Structural Vulnerability in Reproductive Health}
\label{subsec:misinformation}

Unfortunately, when seeking information from multiple channels, sometimes misinformation is present. This is particularly true in reproductive health, which is a highly vulnerable domain, historically and currently, due to clinical complexity, social bias, and historical and ongoing inequities~\cite{antoniak2024nlp}. The stakes of misinformation are high, including harmful health decisions and policy consequences~\cite{john2025online}. Prior work documents the wide prevalence of misinformation online~\cite{suarez2021prevalence}, the impacts of misinformation like complacency, polarization, and hesitancy~\cite{wu2019misinformation,budak2024misunderstanding}, and how clinical systems can be slow to adopt the latest ACOG guidelines, 10 years later~\cite{kozhimannil2017uptake}. In response, organizations such as the CDC (Centers for Disease Control and Prevention) and ACOG (American College of Obstetricians and Gynecologists) emphasize accurate and timely information as crucial for managing pregnancy and pregnancy-related complications~\cite{cdcprevention,acogredesign}.

Patients may seek alternative channels due to access constraints and institutional frictions, opening them up to potential misinformation.  These information frictions are also shaped by structural disparities in reproductive health~\cite{bailey2017structural,prather2018racism}, with medical mistrust tied to historical bias, such as exclusion of Black patients from obstetric studies and treatment~\cite{montalmant2024racial} (racial bias), socio-economic bias~\cite{veinot2018good}, and religious bias~\cite{padela2017types}. We examine misinformation and how different stakeholders respond to it in their context, intending to identify principles that can help avoid similar harms.

\subsection{Current Clinical Systems and Interventions}
\label{subsec:bridging}

EHRs are generally more reliable in terms of accuracy, supporting secure storage and sharing of patient data, patient-clinician communication, and clinician decision support~\cite{ratwani2017electronic}. While they have been shown to increase patient engagement and access to results and messaging~\cite{reicher2016implementation,kuo2016secure}, the burden of portal communication and EHR has resulted in increased workload on clinicians, creating new bottlenecks~\cite{rotenstein2021differences,adler2020electronic}. 

Recent advances in NLP have enabled systems that retrieve vetted information and answer peripartum questions outside EHR channels. Rosie~\cite{nguyen2024rosie} and Jacaranda Health~\cite{jacaranda} match queries to vetted resources to improve access to factual documents. Antoniak et al.~\cite{antoniak2024nlp} provide principles for ethical use of NLP, with emphasis on context, measurement, and values when applying NLP in maternal health~\cite{antoniak2024nlp}. These principles highlight the need to optimize NLP-based interventions for outcomes that support the entire experience, surfacing concerns on how these tools handle establishing trust and supporting the user throughout their journey in the larger healthcare ecosystem. 

These concerns connect to broader work on fairness, accountability, and transparency in health AI. Health algorithms have reproduced racial bias when administrative proxies stand in for need~\cite{obermeyer2019dissecting}, and health information access is unevenly distributed across social groups and infrastructures~\cite{veinot2018good}. Research on abstraction traps warns against treating fairness as a purely technical property removed from social context~\cite{selbst2019fairness}, while work on algorithmic auditing and trust repair calls for visible accountability when systems fail~\cite{raji2020accountability,tang2024failurecards,pareek2024trust}. Human-AI interaction research similarly shows that trust depends on interaction design and uncertainty communication, not accuracy alone~\cite{buccinca2021trust,liao2022designing}. Together, this work motivates our approach: we probe stakeholder opinions on AI-enabled peripartum interventions in the context of modern language model-based infrastructure and existing socio-technical systems.

\begin{table*}[t]
\centering
\footnotesize
\setlength{\tabcolsep}{0pt}      
\renewcommand{\arraystretch}{1.15}
\captionsetup[subtable]{justification=centering,singlelinecheck=false,skip=2pt}
\caption{Overview of participants.}
\label{tab:focus-overview}
\begin{tabular}{@{}p{0.49\textwidth}@{\hspace{0.02\textwidth}}p{0.49\textwidth}@{}}

\begin{subtable}[t]{\linewidth}\centering\vspace{0pt}
\begin{tabularx}{\linewidth}{@{} c C c @{}}
\toprule
Focus group & Stakeholder group & \# Participants \\
\midrule
1 & \bluebox{Health worker} & 8 \\
2 & \purplebox{Clinical team} & 5 \\
3 & \makecell[c]{\greenbox{Birthing people (health)}} & 4 \\
4 & \makecell[c]{\greenbox{Birthing people (non-health)}} & 7 \\
\bottomrule
\end{tabularx}
\caption{Focus groups by stakeholder and number of participants.}
\label{tab:stakeholder_distribution}
\end{subtable}
&
\begin{subtable}[t]{\linewidth}\centering\vspace{0pt}
\begin{tabularx}{\linewidth}{@{} c C c @{}}
\toprule
\# & Demographics & Count \\
\midrule
1 & \purplebox{OB/GYN} & 1 \\
2 & \purplebox{CNM} & 4 \\
3 & \purplebox{Nurse} & 2 \\
4 & \bluebox{Doula} & 6 \\
5 & \bluebox{Social Worker} & 1 \\
6 & \bluebox{Community Health Worker} & 3 \\
\bottomrule
\end{tabularx}
\caption{Expertise in the \purplebox{clinical} and \bluebox{health worker} groups (participants may have multiple roles).}
\label{tab:provider-roles}
\end{subtable}
\\[0.9em]

\begin{subtable}[t]{\linewidth}\centering\vspace{0pt}
\begin{tabularx}{\linewidth}{@{} c C c @{}}
\toprule
\# & Demographics & Count \\
\midrule
1 & Currently pregnant & 5 \\
2 & Postpartum (<1 year) & 3 \\
3 & Postpartum (1--2 years) & 1 \\
4 & Postpartum (2--3 years) & 2 \\
\bottomrule
\end{tabularx}
\caption{Latest pregnancy/postpartum status of birthing participants.}
\label{tab:demographics-status}
\end{subtable}
&
\begin{subtable}[t]{\linewidth}\centering\vspace{0pt}
\begin{tabularx}{\linewidth}{@{} c C c @{}}
\toprule
\# & Distance & Count \\
\midrule
1 & < 10 miles & 10 \\
2 & 10-50 miles & 9 \\
3 & 50-100 miles & 5 \\
\bottomrule
\end{tabularx}
\caption{Distance from focus health institution (participants from 7 distinct counties).}
\label{tab:distance}
\end{subtable}
\end{tabular}
\end{table*}

\begin{table*}[t]
\centering
\small
\setlength{\tabcolsep}{5pt}
\renewcommand{\arraystretch}{1.25}
\caption{Four-part interview guide structured by topic discussed (in the same order) and how it was presented to various stakeholders in the stakeholder-based focus groups.}
\label{tab:interview_guide}

\begin{tabularx}{\textwidth}{@{}p{0.22\textwidth}YYY@{}}
\toprule
\makecell[l]{\textbf{Topic($\downarrow)$ / Stakeholders ($\rightarrow$)}} &
\makecell[l]{\textbf{Birthing people}} &
\makecell[l]{\textbf{Health workers}} &
\makecell[l]{\textbf{Clinicians}} \\
\midrule

\textbf{Information sources} &
Where do you go when you have questions about pregnancy and related topics?

Walk me through a question that led you to seek answers through these sources. &
When your clients/care recipients have questions about pregnancy and related topics, where do they get information? &
When your patients have questions about pregnancy and related topics, where do they get information? \\
\addlinespace[0.6em]

\textbf{Clinical systems} &
How did you utilize the patient portal? What do you think about:
\tightenum{
\item What works well?
\item What does not work well?
\item What would make this more effective?
} &
How do you utilize a messaging system? What do you think about:
\tightenum{
\item What works well?
\item What does not work well?
\item What would make this more effective?
} &
How did you utilize the patient portal? What do you think about:
\tightenum{
\item What works well?
\item What does not work well?
\item What would make this more effective?
} \\
\addlinespace[0.6em]

\textbf{Misinformation} &
\tightenum{
\item How has misinformation affected you?
\item How did you verify potential misinformation?
} &
\tightenum{
\item How has misinformation affected your care of birthing people?
\item What were common sources and how do you combat misinformation?
} &
\tightenum{
\item How has misinformation affected your clinical care of birthing people?
\item How do you combat misinformation?
} \\
\addlinespace[0.6em]

\textbf{AI-powered FAQ tool} &
\tightenum{
\item What would make you want to use a tool like this?
\item What would deter you from using a tool like this?
\item Which factors are most important to address when developing this tool?
\item What could keep this tool from working well?
} &
\tightenum{
\item What would make you want to use/refer others to a tool like this?
\item What would deter you from using/referring others to a tool like this?
\item Which factors are most important to address when developing this tool?
\item What could keep this tool from working well?
} &
\tightenum{
\item What would make you want to use/refer others to a tool like this?
\item What would deter you from using/referring others to a tool like this?
\item Which factors are most important to address when developing this tool?
\item What could keep this tool from working well?
} \\
\bottomrule
\end{tabularx}
\end{table*}

\section{Methods}
We now describe the decisions, processes, and actions undertaken in conducting the study. We preface this with the note that our study was reviewed and approved by the Institutional Review Board (IRB) of the authors' institution.

\subsection{Study Design}\label{subsec:study_design}
\paragraph{Stakeholders:} We considered three groups of stakeholders who are critically important in the care and information seeking experience of birthing people \cite{backes2020maternal}: \greenbox{Birthing people}, \purplebox{Clinicians}, and \bluebox{Health workers}. %

\paragraph{Focus groups:} We conducted four separate, synchronous focus groups with a total of \totalparticipants participants over Zoom\footnote{\url{https://zoom.us}}. Our rationale behind choosing synchronous focus group methodology was \textbf{(a)} to enable discussion based on shared and lived experiences and \textbf{(b)} based on recommendations from previous works on studies in health involving sensitive topics \cite{reisner2018sensitive,augustaitis2021online}. Following the focus group framework in \cite{reisner2018sensitive}, we aimed to include six to eight participants per focus group. In each of the focus groups, participants had shared expertise -- \purplebox{clinical experience}, \bluebox{health work experience}, or \greenbox{lived experience}, and they were appropriately prompted to draw their answers from these experiences at the beginning of the session. During the focus group sessions, some participants with clinical and health experience also drew some observations from their personal birthing experiences whenever applicable. 
The focus group structure was predetermined and comprised of four parts. It was moderated by one of the authors with experience in qualitative studies (JE), as well as reproductive health, community health, and social work, with at least one author in attendance to answer any queries or concerns the participants had outside of the discussions (VB and/or EBK). The participants were prompted for each of the four parts with a design probe in the last part. 
\paragraph{Recruitment:} The recruitment for the focus group participation was advertised online through various Facebook groups and email listservs with people from different counties up to 100 miles from Michigan Medicine, a leading health institution located in the United States. Tables~\ref{tab:stakeholder_distribution},~\ref{tab:provider-roles}, and~\ref{tab:demographics-status} detail different attributes of our stakeholders. We specifically chose stakeholders within this radius to obtain a complete perspective from people who work with similar local healthcare dynamics alongside the focus institution. The participants were invited for an hour and a half virtual session and were provided compensation of \$75\footnote{The hourly rate was determined based on the National Health Council's fair-market value calculator \url{https://nationalhealthcouncil.org/fair-market-value-calculator/} at the time of the study for participants with lived experience.}. We aimed to include birthing people who have given birth within the last three years or are currently pregnant to account for recency and familiarity with the clinical care system and the digital landscape. Furthermore, the focus health institution reports 80\% of their patients residing within 50 miles, approximately matching our recruitment efforts\footnote{\url{https://medresearch.umich.edu/office-research/about-office-research/our-units/clinical-trials-support-office/michigan-medicine-site-profile}} (see Table \ref{tab:distance}).\phantomsection\label{par:compensation}

\begin{figure*}
  \centering
  \includegraphics[width=0.9\linewidth]{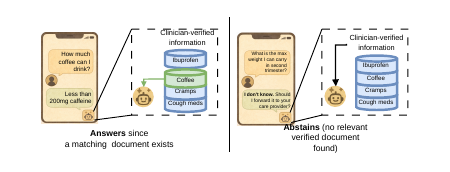}
  \caption{Illustration of the probe depicting an AI-powered FAQ application that matches a user query to clinician-verified documents associated with the query. The same illustration was displayed to all stakeholders during the study.}
  \label{fig:app}
  \Description{Chat interface of an AI-powered application. One phone screen showing a chat where the AI provides answers from a database with verified factual documents and the other phone screen showing a chat where the AI abstains from answering a question and suggests forwarding it to the care provider since there are no matching documents to the query.}
\end{figure*}

\paragraph{Interview Design:}\label{par:interview_design} We designed a four-part structure consistent across all stakeholder groups, with questions tailored to each group's experience and expertise. Table~\ref{tab:interview_guide} details the full structure. We kept the four topics constant so that we could compare responses across groups, but tailored the wording to each group's role: birthing people were asked to respond based on their lived experience, while clinicians and health workers were asked to respond based on their professional roles. We used the same AI-powered FAQ probe across groups to hold the interaction constant and surface differences in governance expectations. We summarize the topics and the rationale behind choosing them:

\begin{itemize}
    \item[\textbf{1.}] \textbf{Information sources: } Gather various sources being used; understand the reason behind choosing sources.
    \item[\textbf{2.}] \textbf{Patient portal: } Understand the role of the messaging functionality in EHRs. For brevity, we hereby refer to this as the patient portal.
    \item[\textbf{3.}] \textbf{Misinformation: } Understand the perspectives on misinformation they received and how they sought to mitigate the effects.
    \item[\textbf{4.}] \textbf{AI-powered FAQ tool:} Showcase an AI-powered tool (Figure \ref{fig:app}) that matches questions to fact-checked documents from clinicians, if available, or abstains from answering. This is used as a probe to elicit stakeholders' critiques and preferences for improving it. We refer readers to Figure \ref{fig:design_probe_abstain} for the actual figures shown during the focus groups. We keep the probe simple with one clear functionality (providing information or abstaining) and one governance prompt (should I forward it to your clinician) to explore open-ended preferences and alternatives during the discussion, similar to principles described in~\cite{hutchinson2003technology}.
\end{itemize}

Our study design is similar to recent work on NLP for maternal healthcare~\cite{antoniak2024nlp}, which similarly argues that AI systems in maternal health should be assessed in relation to the broader care experience rather than only technical accuracy. In this work, we use the design probes and synchronous focus groups to surface how different stakeholders evaluate these systems in a shared, local context and the governance values they expect.

\phantomsection\label{par:pos_statement}
\subsection{Analysis Methodology}

Due to the qualitative nature of our focus group and data, we adopted an inductive open coding approach~\cite{charmaz2006constructing} to analyze the transcripts. First, the third author obtained the transcripts created by Zoom, manually corrected and verified them, and segmented the material by stakeholder group and topic. We then used inductive open coding on the corrected transcripts to generate an initial code set for each focus group. The final author conducted the first round of coding across all focus group data, followed by other authors. %
We then merged overlapping codes across groups, refined the codebook through focused coding, and resolved disagreements through repeated author discussions until reaching consensus on the final thematic structure. We note our study and analysis methodology are similar to other works researching information seeking~\cite{augustaitis2021online}. We preface the results with a belief that AI-based technologies have the potential to overcome gaps in care, and approach our analysis to uncover such gaps.

\section{Results}
\label{sec:results}
\begin{table}[t]
\centering
\small
\setlength{\tabcolsep}{3pt}
\renewcommand{\arraystretch}{1.25}
\begin{tabularx}{\columnwidth}{@{}p{0.25\columnwidth}p{0.35\columnwidth}p{0.35\columnwidth}@{}}
\toprule
\textbf{Theme} & \textbf{Core Finding} & \textbf{Design Implication} \\
\midrule
\textbf{1. Supporting social sensemaking} & 
Seeking reassurance and ``people like me,'' not just facts & 
Support-first patterns; opt-in peer narratives; mistrust-aware design \\
\addlinespace[2pt]
\textbf{2. Verification Pluralism} & 
Divergent trust postures shape what counts as ``credible'' & 
Multiple checking pathways; self-advocacy toolkits; comparison scaffolds \\
\addlinespace[2pt]
\textbf{3. Inspectable Governance} & 
Trustworthiness must be demonstrable, not asserted & 
Provenance metadata; auditable feedback; service-level commitments \\
\addlinespace[2pt]
\textbf{4. Ecosystem complementarity} & 
Tools must complement workflows, not burden them & 
Resource routing; lightweight handoffs; actionable responsiveness scaffolds; \\
\bottomrule
\end{tabularx}
\caption{Summary of themes with core findings and design implications.}
\label{tab:themes-compact}
\end{table}

We uncover four themes that surface how peripartum information seeking is shaped by mistrust and institutional frictions, and how stakeholders emphasize the need for ecosystem complementarity and strong accountability of AI-powered interventions. \textbf{Across themes, stakeholders consistently reframed ``good information'' from being merely correct to being actionable, context-aware, and safely integrated into care pathways}. In particular, themes 1 and 2 mainly stem from general information-seeking experiences, while Themes 3 and 4 stem from discussions of the AI probe and how it fits within existing care infrastructure. For each theme, we outline (i) key observations, (ii) sub-findings with supporting quotes, (iii) design artifacts interpreted from participant needs, and (iv) prior literature that has also shown the importance of these design artifacts (when relevant).

\subsection{Theme 1: Social Sensemaking and Historical Contexts are Key to Trustworthiness}
\label{subsec:theme1_social}
Our first theme touches on the idea of context within ``good information.'' All stakeholders uniformly noted that birthing people do not simply seek facts, but also seek reassurance, normalization, and stories from other people \q{like me.}  Some participants attribute this to social sensemaking, while others describe the influence of past interactional experiences. We identify two interconnected sub-findings: first, that experience-seeking from peers serves a distinct informational function beyond fact-finding; and second, that this reliance on peer networks is often amplified by negative experiences within clinical care systems. These sub-findings have design consequences: AI  systems must also account for the affective and relational nature of information seeking.
\paragraph{1.1 Experience-seeking is a primary function, not a failure mode.}
Participants described turning to social media, community groups, and trusted family members even while recognizing these spaces contain misinformation. This suggests that peer narratives must be included as a legitimate information function and grounding function that AI-mediated support must accommodate to help support the affective needs of birthing people.

\begin{participantquoteBP}
\q{\textit{... I also found Reddit was useful for specific, obscure questions where there are surprisingly a lot of people that have answers or similar experiences.}}
\quoteattrib{Birthing person 4}
\end{participantquoteBP}

\paragraph{1.2 Mistrust is produced by interactional harms and unmet care expectations.}
Health workers and birthing people described feeling misheard, facing slow turnaround, and encountering interactional friction in clinical care, which motivates reliance on faster or personally trusted non-clinical sources.

\begin{participantquoteHW}
    \q{\textit{(birthing people) are given the hurry up and get this like figured out and done with right instead of it really being given the opportunity to flow in a way that they (birthing people) want}}
    \quoteattrib{Health worker 5}
\end{participantquoteHW}

\begin{participantquoteBP}
    \q{{\textit{Is the tool taking into effect other social detriments of health, social determinants of health. We know that things are different for Black and Brown folks than they are for white folks. We know that things are different for people with lower socio-economic backgrounds than folks who don't know that things are different for people who have a higher education than if people have a you know, an ethnic sounding name on their chart that they get different treatment automatically}}}
    \quoteattrib{Birthing person 9}
\end{participantquoteBP}

This account reflects well-documented findings where Black and Brown patients receive different treatment due to structural inequities in obstetric care~\cite{owens2017medical,crear2021social}, termed obstetric racism~\cite{davis2019obstetric} and subsequently extend to other disparities like religious~\cite{padela2017types}, socio-economic~\cite{veinot2018good}, and many more. These experiences of mistrust shape how birthing people evaluate and seek information, making mistrust-aware design not just a usability preference, but a response to harm rooted in history and still ongoing.

\recommendation{
\begin{enumerate}
    \item \textbf{Support-first interaction patterns:} Acknowledge, normalize, and calibrate uncertainty before offering resources.
    \item \textbf{Opt-in peer narratives:} Lived experiences that offer similar factual accuracy should similarly be presented if it is the preferred mode to receive information.
    \item \textbf{Mistrust-aware information design:} Recognize that information and recommendations regarding historically sensitive topics require historical acknowledgment and providing accurate context on what information (personal, surrounding, etc.) and clinical guidelines were used to make the recommendation.
\end{enumerate}
}
\paragraph{Prior literature.} These design artifacts echo some design choices made by previous works like \cite{progga2023understanding,progga2024large} which show how storytelling and social support dynamics shape what information feels credible and actionable.
\subsection{Theme 2: Applications Should Adopt Pluralistic Trust and Verification Practices for the User}
\label{subsec:theme2_verification}
Safety in information seeking was another key, context-dependent need. Stakeholders diverged on what makes any advice or information \q{credible.} We use \textit{pluralism} here to acknowledge that how people trust and verify information varies across individuals and communities. For some topics like vaccine safety, a clear clinical consensus exists, but different approaches may be needed to communicate it effectively to different audiences. For other topics like dietary choices during pregnancy, clinical guidance may be limited or culturally variable, requiring different verification strategies.
\paragraph{2.1 Supporting Divergent Verification Mechanisms.}
Some birthing people treat clinicians as the gold standard, while others view clinical systems as slow to update practices or misaligned with their needs. Health workers emphasized verification through patient education and structured decision strategies such as BRAIN (Benefits, Risks, Alternatives, Intuition, and Nothing), highlighting that verification is frequently performed through dialogue rather than individual, independent fact-checking.

\begin{participantquoteBPpair}
  \panelcaption{Trust in the clinical system}
  \q{\textit{... similar to what others have said. I trust my provider. I trust that they're relying on the most up-to-date medical information through testing research studies}}\quoteattrib{Birthing person 4}
\tcblower
  \panelcaption{Distrust in the clinical system}
  \q{\textit{... it takes a very, very long time for hospital culture to change, so even if there are new evidence-based practices getting the hospital staff on board with is not always easy}} \quoteattrib{Birthing person 9}
\end{participantquoteBPpair}

\paragraph{2.2 Trust calibration is a safety feature.}
Across groups, participants emphasized that systems should avoid individualized symptomatic interpretation and should handle ambiguity responsibly. This implies that safe support requires active trust calibration, including principled abstention, uncertainty communication, and clear handoffs. For instance, health workers described safe support as helping birthing people judge urgency and next steps without actually replacing clinical judgment.

\begin{participantquoteHW}
    \q{\textit{(BRAIN tool) has to do with the mom, being able to ask specific questions, to get more information on whether or not she can. Does something need to be done right now? Is this something where I can wait. Are there other alternatives that I can use? Or can I just do nothing, and just wait for my gut to tell me what to do and listen to my body}}
    \quoteattrib{Health worker 8}
\end{participantquoteHW}

\recommendation{
\begin{enumerate}
    \item \textbf{Verification pluralism:} Offer multiple checking pathways like guideline references, clinician consultation prompts, \q{what would change this recommendation} explanations, and public discussions from trusted members of the community wherever applicable.
    \item \textbf{Self-advocacy toolkits:} Guided patient-centric workflows like BRAIN to generate tailored questions to support decision-making. This artifact is directly derived from recommendations during the focus group session with health workers.
    \item \textbf{Supporting multimedia sources:} Enabling users to share sources like blogs, social media posts, etc., and receiving structured agreements and disagreements to understand the factual stance behind the information.
    \item \textbf{Calibrated support / abstain provision} Provide users to opt out based on health risk, uncertainty in answers, and complexity of symptomatic interpretations before presenting the relevant information based on their comfort. In discussions around the capabilities of the system to abstain, varying preferences around a strong need for abstention to \textit{providing something in the meantime} led us to this artifact.
\end{enumerate}
}

\paragraph{Prior literature.} Prior works in human-AI interaction\cite{buccinca2021trust} and trustworthy AI design\cite{liao2022designing} also encourage calibrated support via cognitive forcing functions to reduce overreliance and user-facing cues, respectively.
\subsection{Theme 3: Provenance, Recourse, and Privacy are Central to Application Credibility}
\label{subsec:theme3_governance}

With the AI design probe, we further examined safety and context in information seeking with AI. Here, we describe the governance infrastructure that stakeholders discussed and identified as important for the credibility of AI-powered tools. In discussions surrounding the design probe (Figure \ref{fig:app}), questions like \q{\textit{who validated this?}}, \q{\textit{when was this validated?}}, and \q{\textit{what happens if it is wrong?}} were common across the groups. These questions reveal that in high-stakes domains like reproductive health, trustworthiness must be demonstrable through artifacts and transparent governance of the information and the AI being deployed.
\paragraph{3.1 Information provenance includes who, authority, and history}
Participants emphasized that usefulness depends on knowing who validated information, recency, and validity in the current age, and metadata indicating validating authorities, timing, and history of changes.
\begin{participantquoteCL}
    \q{\textit{Who would be managing this and keeping it up to date? The tool gets developed, but then how do we remain up to date?}} 
    \quoteattrib{Clinician 1}
\end{participantquoteCL}
\paragraph{3.2 Accountability requires recourse, escalation, and error reporting.}
Stakeholders requested a clear chain of command and mechanisms for reporting errors (like misalignment with providers), with a need for transparent handling of errors and abstention mechanisms (what to do when the AI refuses to answer) and notification of confirmation (similar to transparency).
\begin{participantquoteBP}
    \q{\textit{...but it (AI) contradicts what your doctor says, because your doctor is still old school}}
    \quoteattrib{Birthing person 9}
\end{participantquoteBP}
\paragraph{3.3 Privacy and data-sharing rules are front-stage requirements.}
Stakeholders requested strict, simple privacy rules governing data sharing across birthing people, clinicians, health workers, and any AI component. Privacy is a crucial question, as it shapes whether users will ask questions at all.
\begin{participantquoteBP}
    \q{\textit{... are there going to be things that flag you as doing something problematic or illegal when you were just trying to get information, and now your doctor is going to use that against you?}}
    \quoteattrib{Birthing person 3}
\end{participantquoteBP}

\recommendation{
\begin{enumerate}
    \item \textbf{Provenance metadata:} Source type, reviewer role, last-reviewed date, last-revised date, and historical context (if any) attached and accessible to every answer.
    \item \textbf{Update sensitivity labels:} Explicitly highlight answers and topics where guidance has changed frequently or is tailored differently to different people. We interpret this as a need to explicitly flag answers and topics where guidance has changed frequently, is actively debated, or is tailored differently to different populations. 
    \item \textbf{Auditable feedback loops:} Ability for users to flag content and receive guarantees on confirmation and corrections. The need for existence and guarantee of auditability comes directly from all stakeholder groups.
    \item \textbf{Service commitments:} Published review timelines and public change logs for high-impact updates.
    \item \textbf{Privacy-first design:} Plain-language summaries at first use, granular consent, and minimal data collection by default. This artifact is derived from some participants mentioning specific fears, such as concerns about being reported to child protective services for certain questions.
\end{enumerate}
}
\paragraph{Prior literature.} The need for auditable systems is acknowledged in previous works \cite{raji2020accountability,tang2024failurecards,pareek2024trust} on requirements of auditable, making failures actionable, and recourse for trust repair after failures, respectively.
\subsection{Theme 4: Interventions Must Complement an Overloaded Ecosystem}
\label{subsec:theme4_ecosystem}
While Themes 1–3 focused on information content, verification, and governance, Theme 4 addresses the requirements for tools to fit existing clinical and community workflows in practice. During the discussions, participants largely supported AI-powered applications (illustrated in Figure \ref{fig:app}) but consistently highlighted the need to complement rather than replace existing workflows. These themes correspond to the tool's abstain features and questions regarding which factors are most important in developing the tool.
\paragraph{4.1 Routing is valuable when it is local, affordable, and human-connected.} Stakeholders supported chatbot-like features that reduce search time and route users to appropriate resources, including encouraging human connection when relevant and feasible, including connecting birthing people to doulas, local resource programs, social workers, and other social groups when relevant. They also called for integration with affordable resources and optional connection to similar others.
\begin{participantquoteCL}
    \q{\textit{Here are some really great resources for you while you wait}}
    \quoteattrib{Health worker 2}
\end{participantquoteCL}
\paragraph{4.2 Workflow-aware handoffs reduce burden and increase safety.}
Participants highlighted the need for systems that complement clinical workflows rather than generate new burdens. The most promising support patterns were not ``answering everything,'' but helping users formulate clinically legible messages, set expectations, and escalate appropriately.
\begin{participantquoteCL}
    \q{\textit{It's important to me that it's an additional tool to the (EHR) portal, not a substitute}}
    \quoteattrib{Clinician 5}
\end{participantquoteCL}
\recommendation{
\begin{enumerate}
    \item \textbf{Resources and Human routing layers:} Map questions to vetted locally actionable resources like doulas, community support, and assistance programs, keeping in mind the affordability and accessibility needs of the user.
    \item \textbf{Principled handoffs:} Consistent with principled abstention artifacts in Theme \ref{subsec:theme2_verification}, support reframing user concerns into structured messages with context and urgency markers for clinical teams while retaining agency with the user.
    \item \textbf{Responsiveness scaffolds:} Explain expected reply times and what to do while waiting, depending on the urgency of the information.
\end{enumerate}
}

\section{Discussion}
\label{sec:discussion}

Like prior literature, we observed a significant influence of disparities on people's obstetric experiences and preferences. This, in turn, significantly influences the role AI could play in this socio-technical system, challenging the assumption that trustworthiness in AI systems can be achieved through model-level interventions only~\cite{selbst2019fairness}. Here, we highlight key considerations for AI in medical systems, particularly peripartum support systems, based on our findings. 

\subsection{Medical Mistrust Stemming from Structural Disparities}\label{subsec:mistrust_discussion}

Participants shared personal accounts and concerns about disparities that should shape any AI deployment in obstetrics. For example, participants shared experiences of differential treatment for Black and Brown patients (Theme 1), a spectrum of trust and distrust in existing medical systems (Theme 2), and fears like data being shared with child protective services (Theme 3), reflecting ongoing structural inequities in obstetric care and its influence on trustworthiness. Subsequently, participants also shared preferences and actions like turning to peer networks and social media (Theme 1), diverging discussions about clinicians being the gold standard (Theme 2), due to poor historical experiences such as rushed treatment or treatment denied.

Design artifacts like mistrust-aware information design (Theme 1), verification pluralism (Theme 2), and privacy-first design (Theme 3) are requirements surfaced as a consequence of the distrust. These would be completely missing if fairness is only treated as a technical property, discussed as the \textit{abstraction trap} in ~\cite{selbst2019fairness}. Our findings suggest that fairness in AI-mediated peripartum support must be understood as a property of the service infrastructure, not just the model, and that this infrastructure must be designed with the specific histories and ongoing conditions of the populations it serves (e.g, health information access is unevenly distributed across social groups and infrastructures~\cite{veinot2018good,padela2017types}). Similar principles are articulated in \cite{metzl2014structural}, where authors propose \q{structural competency} to describe a shift in medical education and practice to influence health outcomes that have structural inequities. Analogously, our findings push for structural competency in the design of AI systems.

\subsection{Towards Transparency and Inspectable Trustworthiness of Systems and Services}

Following discussions on fairness, participants consistently asked service-related questions like ``what sources were used,'' ``who validated this,'' ``how current is it,'' ``what happens when it is wrong,'' and ``who is accountable'' (in Themes 3 and 4). This implies that governance questions of updating the data, handling errors, escalating concerns, and communicating transparently are crucial in any AI design.

Past work on transparency has produced influential artifacts for models and datasets, such as model cards~\cite{mitchell2019model}, data cards~\cite{pushkarna2022data}, and failure cards~\cite{tang2024failurecards}, which helped standardize how developers communicate intended use, limitations, and risks. However, our findings suggest that analogous transparency is needed at the \textit{system and service level} with a paradigm shift from trust-by-design to inspectable trustworthiness.  This can further shift the trustworthiness burden away from the users\cite{10.1145/3442188.3445901}.

Concretely, we recommend extending transparency practices into a \textbf{System or Service Card} for AI-mediated information support. Such an artifact would document: (i) scope boundaries, abstention, and safety policies; (ii) provenance schema and validation roles; (iii) recourse pathways; (iv) privacy and consent defaults, retention, attribution of which data was used, and data-sharing rules; and (v) workflow integration assumptions, including impact on other stakeholders. This system-level transparency creates the foundation for human-AI evaluation that measures service performance and accountability over time, not only model outputs.

\subsection{Implication for Human-AI System Evaluations}\label{subsec:hai}
Human-AI research typically focuses on accuracy and overreliance\cite{bansal2021most,buccinca2021trust,lai2021towards,bondi2022role}. Our themes suggest that such evaluations also need to account for the socio-technical needs of stakeholders, like the context of use of the AI system (e.g., experience-seeking alongside information need (Theme 1)), accommodating divergent trust practices while measuring trust in the system (Theme 2), and so on. This aligns with research showing that trust is shaped by uncertainty cues, documentation, and interaction design, not accuracy alone~\cite{liao2022designing}.

Our results, therefore, point to at least four evaluation targets that current frameworks underemphasize:
(i) \textbf{Abstention quality} (refusals that provide constructive alternatives);
(ii) \textbf{Trust calibration} (users understand scope, uncertainty, and boundaries without overreliance);
(iii) \textbf{Self-advocacy outcomes} (ability to let users generate clearer questions, articulate goals, and navigate care relationships to help retain agency in uncertain conditions);
(iv) \textbf{Affective impact} (reassurance and normalization without undue persuasion). 

These targets, drawn from discussions surrounding the AI design probe, complement accuracy-focused evaluation by measuring whether the system supports the relational and governance functions that participants treated as inseparable from useful information.

\subsection{Generalizability Beyond Peripartum Care}

While our focus is grounded in peripartum information seeking, we argue that the themes and design artifacts emerging from this study offer insights for other AI researchers and practitioners in other high-stakes domains. We outline two takeaways for designing pluralistic AI systems and practitioners in high-stakes domains:

Recent works in LLM alignment have explored value pluralism\cite{sorensen2024roadmap}, which formalizes how responses from any system can have a spectrum of reasonable responses (Overton), steered toward a particular response (Steerable) or the distribution of responses matches a population (Distributional). Such works largely study the responses, but rarely explore the larger governance mechanisms around such systems, like explanations, metadata, and safety that are also required by the users. Our findings suggest that pluralism extends beyond what an AI system says to encompass how users can inspect, verify, and contest that information. These requirements suggest that these capabilities must either be designed as part of the overall system or simply tested as abilities of LLMs.

The design artifacts outlined in Section \ref{sec:results}, like provenance metadata, recourse requirements, and calibrated presentation of responses, are likely transferable to stakeholders in other high-stakes domains. Any domain where (a) users have reasons for varying trust in institutions, (b) misinformation poses harmful risks, and (c) information seeking serves social and emotional functions alongside factual ones may benefit from similar design considerations. These requirements also open up requirements in AI-enabled systems to provide guarantees on the provision of additional provenance, the guarantee of recourse (also referred to as lifelong learning in AI literature), and the ability to selectively abstain depending on user needs and risk tolerance. 

\section{Limitations}

Participants in our study have primarily experienced and worked in healthcare in the United States, primarily in English, and draw their observations while working within the healthcare system in one particular state in the United States. While this helps everyone have a common ground during discussion and improves the reproducibility of our findings, it is difficult to generalize this to other populations. 

Based on our findings, we also acknowledge the difficulties and possible limitations with enacting them. Firstly, developing and maintaining an accountable system incurs a significant overhead, especially in topics that have been historically contentious and need deliberation. Additionally, personalization is deeply subjective and culturally dependent, considering that other factors like distrust, preferences, and empathy are important. We acknowledge that direct optimization for these behaviors can often cause LLMs to display sycophantic behaviors \cite{ranaldi2023large}. 

\section{Ethical Considerations}

Our study procedures and recruitment material were approved by the authors' institution's Institutional Review Board (HUM00265646). During the focus groups, we had additional authors in the Zoom meeting to help address queries and support the participants around these topics. Following \cite{reisner2018sensitive,augustaitis2021online}, the participants discussed both positive and negative experiences with a supportive reception from other participants.

\section*{Disclosure of Use of Generative AI}

Generative Artificial Intelligence tools were used in a limited, assistive capacity during the initial stages of the work and while writing the manuscript. Specifically, we used generative AI to (i) produce draft figures incorporated into promotional flyer materials associated with this work (Figures \ref{fig:flyers}), (ii) perform grammar, language polishing, and feedback of manuscript text, and (iii) generate \LaTeX \texttt{\xspace\textbackslash newcommand} commands for repetitive shorthands in the manuscript. All scientific ideas, study design, analysis, implementation, results, and conclusions are original to the authors. The authors reviewed, verified, and edited all AI-assisted outputs.

\section*{Acknowledgments}

This work was partially supported by funding from Google and the University of Michigan (including the Raoul Wallenberg Institute, E-Health and Artificial Intelligence, and the Center for Academic Innovation). Alex Peahl was also supported by Pulsenmore, NICHD, BCBS of Michigan, MHEF, and Molina Foundation. We also thank all the participants for engaging in discussions and the reviewers for insightful feedback and engagement.

\bibliographystyle{ACM-Reference-Format}
\bibliography{main}

\FloatBarrier
\appendix

\section{Design Probe}

\begin{figure}[H]
    \centering
    \includegraphics[width=0.8\linewidth]{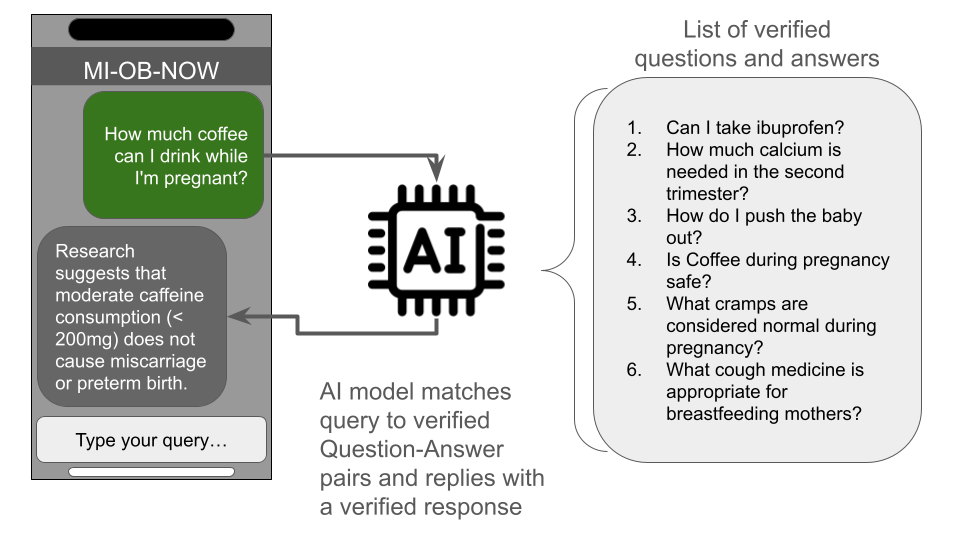}\label{fig:design_probe_actual_pic}
    \caption{Design probe for showing how an AI-powered FAQ app would work.}
    \Description{Chat interface of the AI-powered FAQ application design probe. The phone screen shows a chat where the AI provides answers from a database of factual information documents.}
\end{figure}

\begin{figure}[H]
    \centering
    \includegraphics[width=0.8\linewidth]{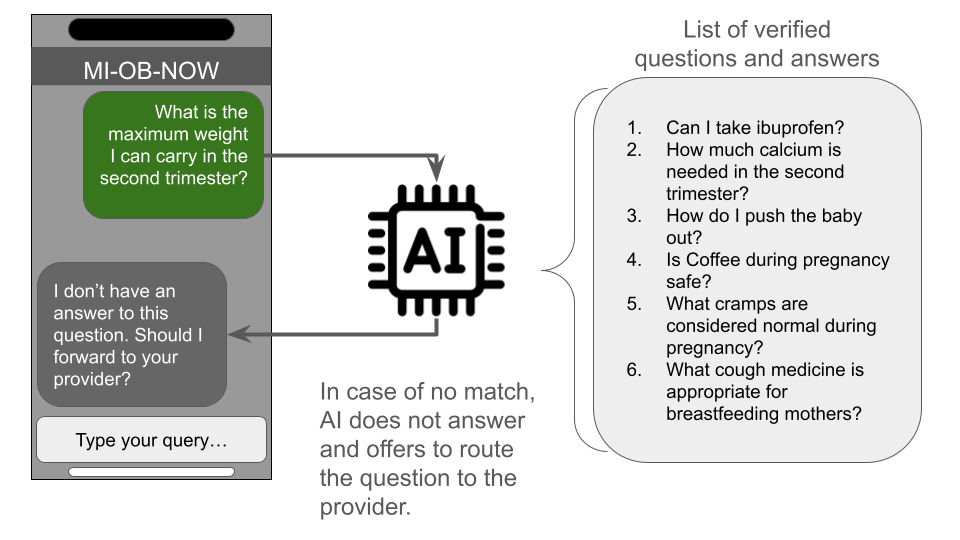}
    \caption{Design probe for showing how an AI-powered FAQ app would abstain when there are no supporting documents available.}\label{fig:design_probe_abstain}
    \Description{Chat interface of the AI-powered FAQ application design probe. The phone screen showing a chat where the AI abstains from answering a question and suggests forwarding it to the care provider since there are no matching documents to the query from a database of factual information documents.}
\end{figure}

\section{Recruitment flyers}

\begin{figure}[H]
  \centering
  \begin{subfigure}[t]{0.49\textwidth}
    \centering
    \includegraphics[width=\linewidth]{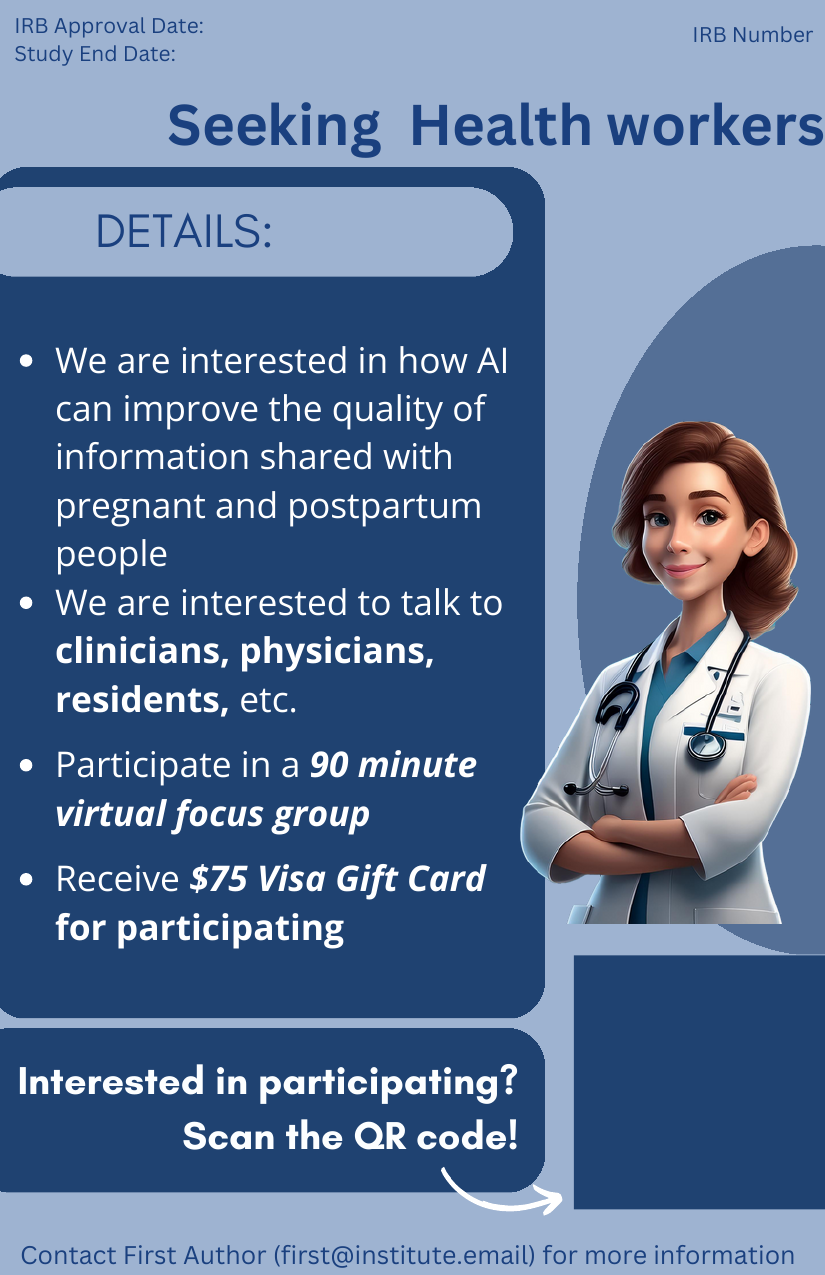}
    \caption{Clinicians}
    \label{fig:clinicians_flyer}
  \end{subfigure}\hfill
  \begin{subfigure}[t]{0.49\textwidth}
    \centering
    \includegraphics[width=\linewidth]{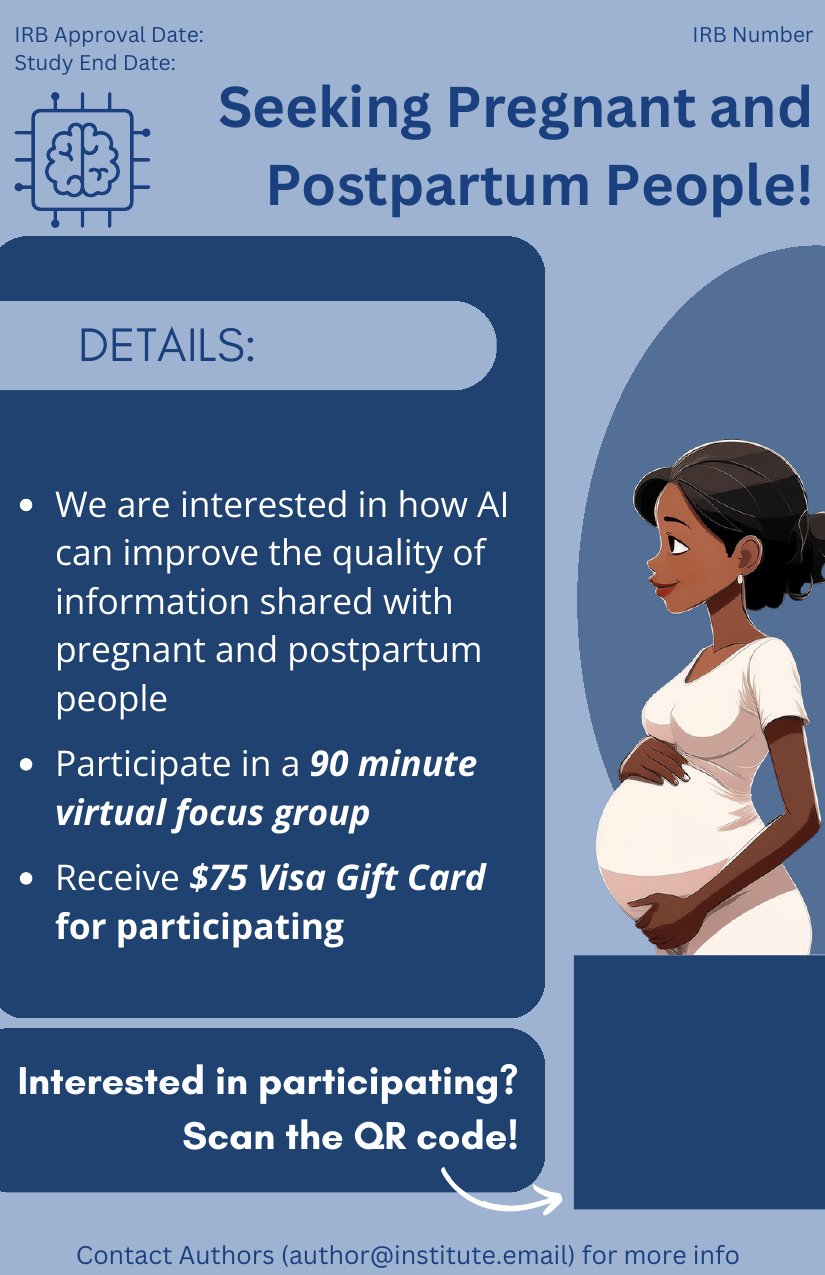}
    \caption{Birthing people}
    \label{fig:patient_flyer}
  \end{subfigure}
  \caption{Recruitment flyers for clinicians and birthing people.}
  \label{fig:flyers}
  \Description{Two recruitment flyers describing the goal of a 90 minute focus groups looking for clinicians (on the left) and birthing people (on the right). The participants are informed about compensation via a \$75 visa gift card and are asked to scan the QR code to register as potential participants.}
\end{figure}

\section{Initial codes}

\begin{longtable}{|c|M{2cm}|M{2.8cm}|M{2.8cm}|M{3cm}|M{2cm}|}
\caption{Focus group codes across three groups with overall synthesis}
\label{tab:focus-groups}\\
\hline
\multicolumn{2}{|c|}{\textbf{Themes \& Sub-themes}} &
\greenbox{\textbf{Birthing people}} &
\bluebox{\textbf{Health workers}} &
\purplebox{\textbf{Clinical team}} &
\takeaway{\textbf{Overall}} \\
\hline
\endfirsthead

\hline
\multicolumn{6}{|l|}{\small\emph{Table \thetable\ (continued)}}\\
\hline
\multicolumn{2}{|c|}{\textbf{Themes \& Sub-themes}} &
\greenbox{\textbf{Birthing people}} &
\bluebox{\textbf{Health workers}} &
\purplebox{\textbf{Clinical team}} &
\takeaway{\textbf{Overall}} \\
\hline
\endhead

\hline
\multicolumn{6}{r|}{\small\emph{Continued on next page}}\\
\endfoot

\hline
\endlastfoot

\multirow{2}{*}{\rotatebox[origin=c]{90}{\textbf{Info. Seeking}}}
& \textbf{Sources} &
Clinicians, Search Engine, Social Media, Trusted People (friends with obstetric/midwife/doula experience), Reputable Sources &
Social Media, Books, Podcasts, Communities (e.g pregnancy center, mom meetups), Trusted People (friends with obstetric/midwife/doula experience) &
Search engines results and summaries, Social Media, Communities, Reputable Sources, Podcasts, and Trusted People (family) &
Social media, Trusted People, Communities \\
\cline{2-6}

& \textbf{Reasons} &
Alleviate anxiety between issue and appointment, Looking for people with similar experience, Difficulties in triage,appointment, ER, Need for Trusting clinician, When online resources are confusing &
Looking for similar experiences, Fear of imperfection, Inadequate clarification or unpleasant clinical experience  &
Looking for similar experiences, trusting people who have gone through these experiences &
Seeking support and similar experiences \\
\hline

\multirow{4}{*}{\rotatebox[origin=c]{90}{\textbf{Patient Portal}}}
& \textbf{Reasons behind usage} &
Track progress, Prescriptions, Information/handouts, After visit summaries, Written documentation on specific questions and answers &
Non-urgent Messaging, Test Results &
Non-urgent messaging, Resources, After-visit summaries, Test Results, Information/Handouts &
Test results, non-urgent messages \\
\cline{2-6}

& \textbf{What works well} &
Test Results, Accessing prescriptions, Scheduling, Answers to specific questions &
maintains boundaries on when to reply to clients &
Shared workload with a team; Flexible for patients and providers; Patient education and resources, Ability for asynchronous messaging &
Test results \\
\cline{2-6}

& \textbf{What does not work well} &
Figuring the right topic and team to ask, Overwhelming to message,  Told to use portal over in person conversation, Slow turnaround time, False sense of accessibility/connection to the provider &
Figuring the right team to ask, Lack of emotional / mental support, Hard to follow up, Slow turnaround time, Unsure about who is answering &
Difficult to navigate the portal, additional overhead of clarifying portal usage, handling conflict between general purpose recommendations vs personalized discharge instructions and test results &
Usability issues, lack of personalization, slow turnaround time, false sense of accessibility \\
\hline

& \textbf{What would make this more effective} &
Including a chatbot with vetted information to reduce clinician burden and improved access for underserved communities,  improved UI/UX with the portal &
Automatic links to appropriate resources during the wait time, ability to request a follow-up conversation (acknowledging already large burden), support for multimedia in the interface  &
Improving boundary setting, Ability to have auto-replies describing response times, increased time for providers to handle portal messages, Improving UI/UX  &
Chatbot-style conversations with resources, auto-replies, better search and elements of personalization \\
\hline

\multirow{3}{*}{\rotatebox[origin=c]{90}{\textbf{Misinformation}}}
& \textbf{How has misinfo. affected you/your practice?} &
Group 1: Trust provider that they are up-to-date with patient history and latest guidelines. Actively seek out clinicians; Group 2: Mistrust provider since hospital guidelines are slow to change, medical research tends to ignore certail demographics, common advice given despite every pregnant being different &
Afraid to seek out resources because of misinformed repercussions (e.g CPS for postpartum depression), providers dictating birth experience potentially incorrectly, fear if baby not hitting standardized developmental milestones &
(emerging) Mistrust in providers, Trusting social media information and partial evidence online, not up-to-dat/ "old school" family members, OB is especially challenging because it's internal/unseen/personal - people have different personal experiences and are experts in those but assume generalization, more fear/anxiety and more competition, differences in impacts on younger vs older patients (maybe less dependent on social media), many of these lead to more time needed for one-on-one discussions which doesn't align with patient load &
Trust in medical system contributes to reliance on clinicians; Distrust in medical system encourages use of other resources \\
\cline{2-6}

& \textbf{Verifying / combatting misinfo} &
Checked source reputation (medical journals or clinics, books, are they selling, is it a blog or social media), check with clinicians (vs get written resources to check clinicians) &
Research for clients, especially to have conversation with provider, strategies for self-advocacy (e.g. BRAIN acronym, ask open ended questions, ask for a moment to make decision), prepare in advance for decisions one might need to make so not caught off guard  &
listening, motivational interviewing, affirmative inquiry, linking back to goals, sliders 1 to 10, referring to existing resources and classes, centering (education, community building, and lower provider burden), would like to build doulas or MAS into site for patient education, would like to share easy-to-find mock conversations instead of existing static/hard-to-find patient education resources &
Self-advocacy and improved communication; verify sources and intentions of information provider \\
\cline{2-6}

\multirow{3}{*}{\rotatebox[origin=c]{90}{\textbf{AI-powered FAQ app}}}
& \textbf{Usefulness} &
useful retrieving current documents and resources, including those from patient portal or medical studies or books; could be nice to connect to others with similar charateristics / experiences / symptoms - whether in portal or directly or social media or support groups; could help to have reminders of questions to ask provider based on q's you ask; ideally should include midwives/doulas not just obs &
could provide an accessible, full circle level of help, esp while chaotic in US &
potential to reduce burden and improve from what we have now, engage with patients and spark questions, answer only generic/group prenatal care questions vs individualized/symptomatic questions, patients might react well to typing in questions vs static faq &
Strong support for usefulness and need \\
\cline{2-6}

& \textbf{What would deter from using / recommending this} &
lack of privacy (could deter from asking qs there), just generative instead of retrieval, while it could empower people could also put people in unsafe situation if at odds with providers &
if it's just a frustrating circular bot chat, questions are still unanswered &
who would manage it; needs to be kept up to date; dangerous to answer symptomatic questions or have patients self-assess risk to provide individualized replies; don't want to redirect people to provider inbox; don't want to replace portal where lots of specific requests are made &
Bad user experience, lack of strong privacy protections and transparency in authenticity \\
\cline{2-6}

& \textbf{Important factors} &
carefully think about validation (who vets it and when), ambiguity, different user preferences about level of detail and risk, politics, avoid minimizing feelings/experiences, keeping it updated, who is building it, whether social determinants of health are included &
will it include holistic perspective in addition to medical (and if so, does that impact patient/provider relationship), who will check messages forwarded to providers, will it support accessibility (multiple languages, visual impairment, literacy), when should it connect to provider directly/automatically vs remain private, can patient have a shortcut or be sent to immediate follow up to provider or resource if urgent (eg mentioning suicide), will it have some emotional support (eg what you're feeling is felt by other people), will it link to research so people can do more research &
could consider individualization in future (eg multiples or singleton, high low risk), but still mostly just standards/non-urgent qs with disclaimers that answers are for routine healthy pregnancies; emphasize cultural inclusion and engagement with an app; give fast answers like google ai with clickable options to learn more; coach towards waiting for next appt or calling if emergency vs messaging provider immediately; don't be a barrier to reaching provider like having to go through this first; only carefully pull from prior clinician answers that could be outdated &
(1) validation - (who vets, how, and how frequently), (2) accounting for ambiguity and user preferences (eg just provide standards (general preference) or personalize based on high or low risk, user preferences around detail and level of risk), (3) include human / cultural / personal / empathetic aspects, (4) include people from range of backgrounds in developing it, (5) consider user friendliness and engagement, (6) what to do when we need to defer (eg coach towards gathering questions for next appointment, directly calling if emergency, how to differentiate) \\
\hline

& \textbf{What could prevent this from working well?} &
tracking might deter questions, might make people scared if not communicated well, research contradicting what old school hospital/dr says, cost, if it fails or is sterile &
if it's paid, or not widely accessible on many devices &
answering symptomatic questions could create more work; people might not use unless it has engagement and sparks questions &
if it's paid or not generally accessible from different devices, if privacy doesn't have a good implementation (eg people are afraid dr could refer them to child protective services for certain questions),  if it's not engaging, if it contradicts what clinicians say, if it fails/answers symptomatic questions \\
\hline

\end{longtable}

\section{Detailed focus group observations} 

In this section, we present key observations based on the four topics discussed during the focus groups. Themes obtained from our analysis across all groups are highlighted as \takeaway{S. Takeaway} (S. stands for findings \textbf{similar across groups}) and \takeawaycontrast{D. Takeaway} (D. stands for findings that \textbf{differed across groups}). We also provide relevant quotes from participants, where the first letter is the stakeholder attribute \greenbox{B} (\greenbox{birthing people}), \bluebox{H} (\bluebox{health worker}) and \purplebox{C} (\purplebox{clinical team}) alongside a number indicating the identifier to distinguish the participants (e.g \greenbox{B5}). Table \ref{tab:results_summary} summarizes the key results by topics of discussion.

\begin{table}[t]
    \centering
    \begin{tabular}{|p{3cm}|p{5cm}|p{5cm}|}
\hline
\textbf{Topic} & \textbf{\takeaway{Similarities}} & \textbf{\takeawaycontrast{Differences}} \\
\hline
Information sources & 
\textbf{(S1.)} Seeking support and similar experiences &
\textbf{(D2)} Complicated clinical care experiences (\greenbox{birthing people}, \bluebox{health workers} differ from \purplebox{clinical team}) \\
\hline
Patient / messaging portals & 
\textbf{(S3)} Lack of patient centricity, \textbf{(S4)} Burden to clinicians, \textbf{(S5)} Difficult user experience, \textbf{(S6)} Improving the patient portal experience &
- \\
\hline
Misinformation & 
- &
\textbf{(D7)} Divide in trust in the clinical system, \textbf{(D8)} Varying focus on verification (\greenbox{birthing people}, \bluebox{health workers} differ from \purplebox{clinical team})\\
\hline
AI-powered FAQ app & 
\textbf{(S9)} Strong support under further conditions and \textbf{(S10)} Eliciting requirements in the application &
- \\
\hline
\end{tabular}

    \caption{Key results from our analysis of each topic. Similarities highlighted are across all three stakeholder groups - \greenbox{birthing people}, \bluebox{health workers}, and \purplebox{clinical team}.}
    \label{tab:results_summary}
\end{table}

\subsection{Information Sources and Reasons Behind the Choice}

\begin{figure}
    \centering
    \includegraphics[width=0.5\linewidth]{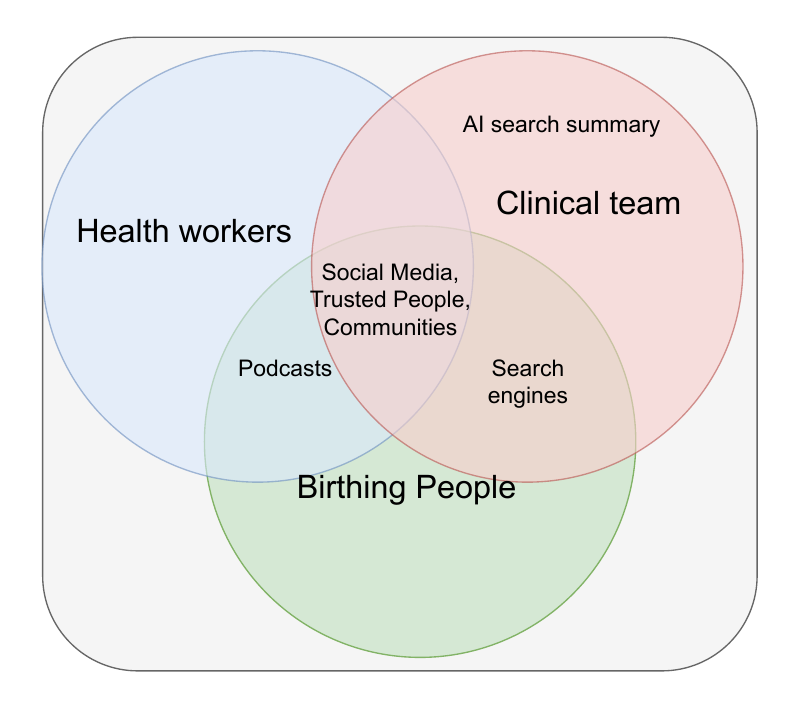}
    \caption{Venn diagram of different sources described by  stakeholders on what birthing people use for information seeking.}
    \Description{
    A Venn diagram illustrating the information sources for birthing people, health workers, and clinical teams. 
    Three overlapping circles represent these groups. Health workers’ circle includes “Podcasts,” the clinical team’s circle includes “Search engines, AI search summary,” and the birthing people’s circle has no standalone label. 
    At the intersection of all three circles are “Social Media, Trusted People, Communities.” 
    The diagram is enclosed in a rounded square background to emphasize the shared context. 
    }
    
    \label{fig:venn_sources}
\end{figure}

We find a variety of sources mentioned by different participants and stakeholders when asked \q{where do you/birthing people go to with questions about pregnancy and related topics?}, as illustrated in Figure \ref{fig:venn_sources}. We also report the queries birthing people report. 
During this discussion, we additionally uncover two key themes surrounding the reasons behind choosing  sources.

\paragraph{Sources} Across all stakeholder groups, participants frequently mentioned social media platforms like TikTok and Instagram, trusted people (friends, family, specially ones with background in clinical care, etc.), podcasts and community groups like Facebook groups and \q{mom-meetups}. Other sources like search engines (e.g., Google), specific websites, and books were also discussed. Figure \ref{fig:venn_sources} illustrates the different sources mentioned by stakeholders. Interestingly, podcasts were only mentioned by birthing people and health workers, and search engines were mentioned only by clinicians and birthing people. The AI-generated summaries now prevalent in search results were only mentioned by clinicians. This may indicate emerging trends, as clinicians and health workers meet many birthing people, and even potential disconnects between different stakeholders.

\paragraph{Queries} Reported queries included seeking information about safe medication, symptom management, complication management, lifestyle choices and interpreting test results. Furthermore, some participants from birthing people focus groups also mentioned using online resources to find evidence for advice they had received. 

\paragraph{Themes} We next highlight two key themes from these information seeking discussions and analyses. \\

\takeaway{(S1) Seeking support and similar experiences:} A key reason that was frequently stated across all the groups was the sense of birthing people to find others with similar experiences and support, while being fully aware of the presence of misinformation in online platforms and the dangers of online information.

 \q{\textit{... constantly looking on <search engine>, which I know you are not supposed to about progress and complications. I also found Reddit was useful for specific, obscure questions where there are surprisingly a lot of people that have answers or similar experiences.}}-\greenbox{B4}\\

\takeawaycontrast{(D2) Complicated clinical care experiences:} In analyzing the reasons behind choosing specific information sources, we uncover the complicated relationship that exists between birthing people and the clinical system, which arises again in discussions of misinformation (\takeawaycontrast{(D7)}, \takeawaycontrast{(D8)}). Health workers and birthing people expressed complicated clinician-patient experiences as a  key factor in seeking information from non-clinical care resources. Complicated experiences include instances where the birthing people have felt misheard or otherwise had negative experiences with the clinical care system, etc. In these cases, they often reverted to familiar, personally trusted sources. Furthermore, other experiences surrounding slow turnaround and waiting on  patient messaging, triage, between appointments, etc. was another reason to why these alternate sources were preferred. Two quotes from different stakeholders illustrate the complexity of their individual experiences:

\begin{figure}[htbp]
  \centering

  \begin{subfigure}[t]{0.49\textwidth}
    \centering
    \begin{displayquote}[— \greenbox{B2}] \q{\textit{... So any questions that are along that vein (participant's condition) the sources that I go to [are] our clinicians. ... the sources I look for are there a clinician involved in this or not? Whether it's a book, it's an article.}}
    \end{displayquote}
    \caption{Rely on clinicians}
  \end{subfigure}
  \hfill
  \begin{subfigure}[t]{0.49\textwidth}
    \centering
    \begin{displayquote}[— \bluebox{H5}]
      \q{\textit{(birthing people) are given the hurry up and get this like figured out and done with right instead of it really being given the opportunity to flow in a way that they (birthing people) want}}
    \end{displayquote}
    \caption{Conflict with  clinicians}
  \end{subfigure}
\end{figure}

\subsection{Patient Portals: Utility and Grievances}

\paragraph{Utility} We find that test results -- notification, interpretation,  discussion, and follow-throughs -- play a key role in the utility of the patient messaging portal, acknowledged by most stakeholders. Clinicians also recognized the importance of these portals in patient education and providing after-visit summaries, all in one place, and appreciate the ability to potentially share the messaging workload with a larger team. 

\paragraph{Themes} We next highlight four key themes emerging from these portal discussions and analyses. The unifying observation with these portals is their lack of usability, across all stakeholders. We also include reported ideas for improving the portal experience. \\

\takeaway{(S3) Lack of patient centricity} We find that both birthing people in their experience and health workers in their practice have observed the difficulty in finding the right team and right topic to send a message to. There is also a potential lack of personalization, where birthing people face a false sense of accessibility and connection, while lacking personal and emotional support over text. Personalization is challenging for clinicians as well, who acknowledge that personalized peripartum advice can conflict with general guidelines that may be included in clinician-centered time-saving messaging tools.

\begin{displayquote}[-\greenbox{B9}]
    \q{\textit{there's like this, kind of a false sense of accessibility with the portal. We feel like we can send a message 24/7. But we're not going to get a response. 24/7.}}
\end{displayquote}

\takeaway{(S4) Burden to clinicians} Clinicians collectively acknowledge the burden and burnout from handling messaging on the patient portal, where the built-in time for them to handle these messages is insufficient. Birthing people wonder how much to engage via the portal, given the  burden their questions could create and even the potential costs associated with sending messages in some health systems.

\begin{displayquote}[\purplebox{C4}]
   \q{\textit{It is also a source of challenge for physician wellness and provider wellness.}}
\end{displayquote}

\takeaway{(S5) Difficult user experience} All stakeholders collectively state the lack of user centricity in the design and workflows involved in these portals, both from the patient and clinician perspectives. For instance, birthing people report having to search for information they need, as it is buried in  long reports across various visits. Additionally, clinicians report having to teach birthing people how to navigate to the right resources within the portal, and difficulty using features meant to save time, like smart phrases that auto-complete common messaging. 

\begin{displayquote}[-\greenbox{B3}]
    \q{\textit{but there's like an after visit Summary, but some of them have been like 14 pages long, and if I want to go back like someone else mentioned, like what medications are safe in pregnancy, if I want to remember which medications that they say to take when I get a migraine, I don't have to have to remember, not only like where in the after visit? Summary. \textbf{But which visit did we talk about in April, or did we talk about that in May? And so for me it is a lot faster to just Google it than to try to go through pages}. And I have to download. I have to log into the portal.}}
\end{displayquote}

\takeaway{(S6) Improving the patient portal experience} We find that all stakeholders collectively report integrating a chatbot-like functionality that can provide auto-replies with resources, expected response time and personalized messages (including multimedia like audio and video) on behalf of the clinical team and reduce the search time for all  stakeholders. Interestingly, we note that conversations around the chatbot were often  organically brought up before the moderator mentioned an AI-powered FAQ tool.

\begin{displayquote}[-\bluebox{H3}]
    \q{\textit{I would love to see the ability for clients or patients and their providers to audibly record instead of typing things, because sometimes typing is overwhelming, and it doesn't have the tonality and the inflection, especially if there's somebody who English is a second language}}
\end{displayquote}

\subsection{Stakeholders Tackling Misinformation}

The discussions surrounding misinformation revealed many issues surrounding trust and divides that exist between the clinical care system and birthing people.

\takeawaycontrast{(D7) Divide in trust in the clinical system} In the discussion surrounding impact of misinformation, participants describe their personal experiences with misinformation, revealing a divide in their process of verification. Some participants trust their clinical care providers to be the gold standard of trustworthy and up-to-date information, %
while other participants have distrust, reporting clinicians as a potential source of misinformation, especially due to hospitals being slow to update their policies and current research not being truly reflective of specific demographics. Clinicians also reported similar experiences where patients with mistrust in the system tend to prefer other sources of information like social media.

\begin{figure}[htbp]
  \centering

  \begin{subfigure}[t]{0.42\textwidth}
    \centering
    \begin{displayquote}[— \greenbox{B4}] \q{\textit{... similar to what others have said. I trust my provider. I trust that they're relying on the most up to date medical information through testing research studies}}
    \end{displayquote}
    \caption{Trust in the clinical system}
  \end{subfigure}
  \hfill
  \begin{subfigure}[t]{0.57\textwidth}
    \centering
    \begin{displayquote}[— \greenbox{B9}]
      \q{\textit{it takes a very, very long time for hospital culture to change, so even if there are new evidence-based practices getting the hospital staff on board with is not always easy. You know there's no benefit to remaining pregnant past 39 weeks. except for all the side effects and risks of a potential induction like they didn't bring that up at all.}}
    \end{displayquote}
    \caption{Distrust in the clinical system}
  \end{subfigure}
\end{figure}

\takeawaycontrast{(D8) Varying focus on verification} The divide of stakeholders with both trust and distrust in the clinical care system uncovered different strategies to mitigate misinformation. While birthing people focused on the intentions and sources of information they were receiving, healthworkers and clinicians focused on patient education, open communication and self-advocacy (using techniques like BRAIN (Benefits Risks Alternatives Intuition and what happens Next or if I do Nothing) as the key to combating misinformation. We believe both are motivated by a desire to improve decision-making in an information environment with sources of varying reliability. %

\begin{displayquote}[-\bluebox{H8}]
    \q{\textit{(BRAIN tool) has to do with the mom, being able to ask specific questions, to get more information on whether or not she can. Does something need to be done right now? Is this something where I can wait. Are there other alternatives that I can use? Or can I just do nothing, and just wait for my gut to tell me what to do and listen to my body}}
\end{displayquote}

\subsection{Digital Interventions to Support Information Seeking}

Finally, we present findings on the role of digital interventions (Figure \ref{fig:app}), similar to Rosie \cite{nguyen2024rosie}, in information seeking for pregnancy and postpartum periods. 

\takeaway{(C9) Strong support under further conditions} The participants generally approved of such an intervention, with reasons around reducing burden, increasing access, engaging with patients, retrieving the latest documents and resources, and connecting with others with similar characteristics/experiences/symptoms. However, skepticism and additional requirements were voiced during the discussions on additional features and expectations from such an application, which are described in the next result.

\begin{displayquote}[-\greenbox{B5}]
    \q{\textit{I guess if I want a quick answer I would just go there and do it. Perhaps then I would compare it with whatever I hear on the Internet or from other people.}}
\end{displayquote}

\begin{displayquote}[-\bluebox{H6}]
    \q{\textit{Very excited to see what this does in the community.}}
\end{displayquote}

\takeaway{(C10) Requirements}

Participants suggested requirements to keep in mind for developers, especially related to concerns about potential digital interventions.

\begin{displayquote}[-\purplebox{C1}]
    \q{\textit{I think that would be a helpful tool, my kind of my brain, one of my 1st thoughts would. Who would be managing this and keeping it up to date. So from a function like a long term functional perspective. So the tool gets developed. But then, how do we remain up to date?}}
\end{displayquote}

\end{document}